\documentclass[prb,aps,twocolumn,showpacs,a4paper]{revtex4}
\usepackage{graphicx}
\begin{document}

\title{Transport through a superconductor-interacting normal\\ metal junction: a phenomenological description}
\author{H.T. Man}
\author{T.M. Klapwijk}
\author{A.F. Morpurgo}
\affiliation{Kavli Institute of NanoScience Delft, Faculty of
Applied Sciences, \\Delft University of Technology, Lorentzweg 1,
2628 CJ Delft, The Netherlands}
\date{\today}

\begin{abstract}
We propose a phenomenological description of electronic transport
through a normal metal/superconductor interface of arbitrary
transparency, which accounts for the presence of electron-electron
interaction in the normal metal. The effect of interactions is
included through an energy-dependent transmission probability that
is inserted in the expression for the current-voltage
characteristics of a non-interacting system. This results in a
crossover from the Andreev to the tunneling limit as a function of
the energy at which transport is probed. The proposed description
reproduces qualitatively the results obtained with formally correct
theories as well as experimental observations. In view of its
simplicity, we expect our approach to be of use for the
interpretation of future experiments.
\end{abstract}


\maketitle

Low-energy electronic transport through the interface between a
normal metal (N) and a superconductor (S) can be understood in terms
of normal and Andreev reflection\cite{Andreev}. For systems where
the effect of the Coulomb repulsion between electrons is negligible,
the interplay between these two processes is governed by the
transmission of the NS interface. If the transmission is high (close
to unity) Andreev reflection dominates and the low-energy
conductance of the interface is higher than the conductance measured
at energies above the gap. On the contrary, if the transmission is
low (tunneling regime), Andreev reflection is strongly suppressed
and so is the low-energy conductance of the NS interface. The
crossover from high to low interface transparency is well described
by the theory\cite{BTK} and it is in good agreement with
experimental results on a variety of systems\cite{Blonder}.

When the normal conductor is a low-dimensionality system, however,
electron-electron interactions play an important role and the
situation is considerably more
complex\cite{Vishveshwara,Maslov,Affleck,Lee}. In this case,
electronic transport through the NS interface is not only determined
by the transmission coefficient, but also by the strength of the
interaction. It is understood quite in general that
electron-electron interactions suppress the probability for Andreev
reflection\cite{Oreg}. For interacting systems, however, a simple
and intuitive picture of the interplay between normal and Andreev
reflection does not exist. Nevertheless, such a picture would be
extremely useful for the qualitative interpretation of experiments.

The purpose of this paper is to present a simple phenomenological
description that captures the important aspects of electrical
transport through a NS interface in the presence of repulsive
electron-electron interactions in the normal conductor. To this end,
we consider the simplest possible case of interacting electrons in a
one-dimensional (1D) ballistic conductor connected to a
superconductor with an arbitrary interface transparency. Motivated
by recent theoretical studies on interacting normal mesoscopic
conductors\cite{Matveev,Kindermann,Bagrets}, the effect of
electron-electron interaction is included as an energy-dependent
transmission probability $T(E)$, which we substitute into the
theoretical expression for the current-voltage characteristics of a
system with no electron-electron interaction. We substantiate the
validity of the proposed phenomenological picture by performing
explicit calculations for the case in which $T(E)$ is given by the
expression valid for a barrier in an infinite 1D interacting
conductor\cite{Matveev}. The outcome of these calculations
reproduces qualitatively the trends found in formally correct
theories\cite{Lee}. We also illustrate how different aspects of our
results can explain different aspects of the transport behavior of
carbon nanotubes connected to superconducting
electrodes\cite{Morpurgo}.

\begin{figure}[htb]
  \centering
  \includegraphics[width=8.5cm]{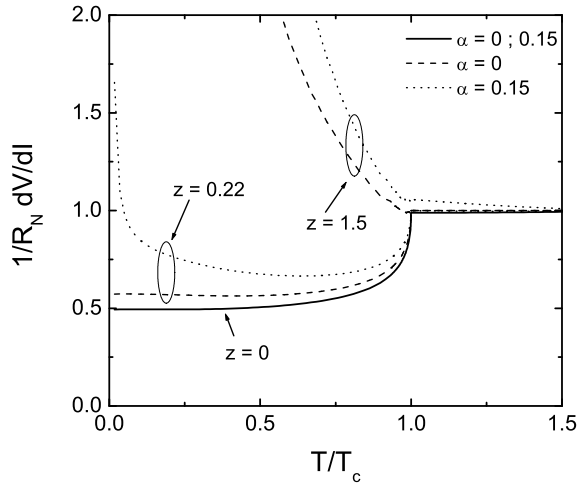}
  \caption{Temperature dependence of the zero-bias differential resistance calculated for a non-interacting
   ($\alpha = 0$, dashed lines) and an interacting ($\alpha = 0.15$,
   dotted lines) 1D conductor connected to a superconductor, for different
    values of the contact transparency ($z=0, 0.22$, and 1.5).
    For $z=0$ (continuous line) the behavior is identical for the interacting
     and the non interacting case. }\label{Temp-Dep}
\end{figure}

Our work is stimulated by the description of electron transport
through a tunnel barrier of arbitrary transmission in a 1D
interacting conductor, which was first proposed by Matveev and
coworkers\cite{Matveev}. Within this picture, the tunnel barrier
induces a Friedel oscillation of the electron density in the
one-dimensional conductor. In the presence of electron-electron
interaction, such a density oscillation creates an electrostatic
potential that contributes to scattering the electrons. Thus, to
find the final transmission coefficient for the tunnel barrier, the
effect of multiple reflections on the barrier and on the Friedel
oscillation needs to be considered. It was shown that the net effect
of the interaction is to introduce an energy dependence in the final
transmission probability, such that the transmission decreases with
decreasing energy.

This description of electron-electron interaction in terms of a
renormalized, energy-dependent transmission coefficient has been
recently extended to deal with more general (i.e., not only one
dimensional) mesoscopic conductors\cite{Kindermann, Bagrets}.
Similarly to the 1D case, it was found that the dominant effect on
low-energy electron transport is due to elastic scattering and it
can be accounted for by an energy renormalization of the
transmission coefficients. At sufficiently low energy, this
renormalization results quite in general in a suppression of the
transmission coefficients with decreasing energy. These conclusions
indicate that, at a qualitative level, accounting for the effects of
electron-electron interaction in terms of a transmission probability
decreasing with decreasing energy is a robust result, insensitive to
the details of the system considered.

\begin{figure}[b]
  \centering
  \includegraphics[width=8.5cm]{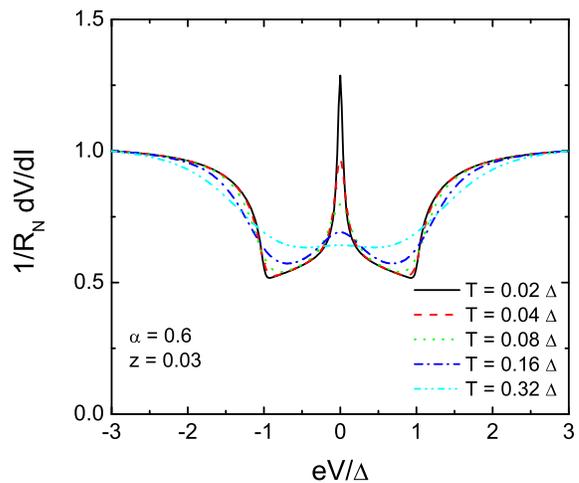}
  \caption{The $dV/dI-V$ characteristic calculated at different
   temperatures below $T_c$, for a high transparency contact
   ($z=0.03$) in the presence of interaction for the same interaction
    strength (corresponding to $\alpha = 0.6$). Note that at higher
    temperature (e.g., $T=0.32\ \Delta$) the $dV/dI-V$ is similar to what is
    observed in the non-interacting case, whereas at lower temperature
    (e.g., $T=0.02\ \Delta$) a zero bias peaks emerge that cannot be
    accounted for in terms of the BTK model. }\label{Diff-Ts}
\end{figure}

For the case of a tunnel barrier with (bare) transmission $T_0$ in
an infinite 1D conductor, the explicit interaction-induced energy
dependence of $T(E)$ reads\cite{Matveev}

\begin{equation}
T(E) = \frac{T_0 \left |\frac{E}{D_0}\right |^{\alpha}}{1 - T_0
\left ( 1 - \left |\frac{E}{D_0}\right|^{\alpha}\right )}
 \label{T_of_E}
\end{equation}

\noindent where $ 0 < \alpha <1$ quantifies the electron-electron
interaction strength ($\alpha = 0$ corresponds to no interactions)
and $D_0$ is a high-energy cut-off determined by the energy
bandwidth of the electronic states. In what follows, we take
$T(E)$ as given by Eq.~\ref{T_of_E} and insert it in the known
expression for the $I-V$ characteristics of a NS interface (with
bare transmission $T_0$) of a non-interacting system.

Formally this procedure is not correct, since the
interaction-induced renormalization of the transmission coefficient
in an infinite 1D conductor differs from that of a tunnel barrier at
a NS interface. This is because the presence of the superconductor
changes the details of the Friedel oscillations and thus the
specific dependence of $T$ on $E$. Nevertheless, calculations
similar to those of Matveev and coworkers performed for a 1D
interacting normal metal/superconductor junction have shown that
also in this case the effect of the interaction is to renormalize
the scattering coefficients\cite{Takane}. More importantly,
Eq.~\ref{T_of_E} captures the trend of an interaction-induced
suppression of the transmission probability that quite in general
accounts for the effect of electron-electron interaction on
transport, as discussed above. Therefore, since our work only aims
at providing a transparent phenomenological description of transport
across a NS interface and not at discussing specific details, the
use of Eq.~\ref{T_of_E} is qualitatively justified\cite{note}.

The second ingredient of our description is the
expression for the current $I$ flowing through a NS interface of
arbitrary transparency as a function of the applied bias $V$. This
was found long ago by Blonder, Tinkham, and Klapwijk (BTK)\cite{BTK}
to be:

\begin{equation}
I = G_0 \int^\infty_{-\infty} dE [f(E-eV)-f(E)][1 + A(E,z) - B(E,z)]
\label{I-BTK}
\end{equation}

\noindent Here $A(E,z)$ and $B(E,z)$ are the energy dependent
Andreev and normal reflection probability, respectively, and $G_0$
is a constant. The parameter $z$ is a constant that quantifies the
amplitude of a delta-like potential barrier at the interface. It is
related to the transmission probability $T_0$ as
\begin{equation}
z^2= \frac{1-T_0}{T_0}\label{z-factor}
\end{equation}
$z = 0$ corresponds to a perfectly transparent interface, resulting
in $A(E,0) =1$ and $B(E,0) =0$ for $E < \Delta$  (Andreev limit); $z
\gg 1$ corresponds to the tunneling limit in which $A(E,z) \ll 1$
and $B(E,z) \simeq 1$.

Whereas in the non-interacting case the $z$ parameter is a constant,
after substituting $T(E)$ for $T_0$ in Eq.~\ref{z-factor} to account
for the presence of interactions $z$ acquires an energy dependence.
We label the energy dependent $z$ parameter as $z_{e-e}$. By direct
substitution we have:
\begin{equation}
z_{e-e}^2 = \left |\frac{E}{D_0} \right |^{-\alpha}\
\frac{1-T_0}{T_0} = \left |\frac{E}{D_0} \right |^{-\alpha} z^2
\label{renorm_z-factor}
\end{equation}
It is clear that now, even for values of $T_0$ close to unity, it is
the energy $E$ which determines whether $z_{e-e}$ is approximately
$0$ or much larger than one. If, for $E < \Delta$, $z_{e-e}$ varies
considerably, electron-electron interactions can induce a crossover
from the Andreev limit to the tunneling limit in an individual
sample, depending on the energy scale on which transport is probed.
This crossover is responsible for the qualitatively different
behavior of systems in which electron-electron interaction plays an
important role, as compared to systems that can be described in
terms of independent electrons. Note, in passing, that if $z = 0$
(i.e., $T_0 = 1$), $z_{e-e}$ is also 0, implying that for a
perfectly ballistic system, electron-electron interactions do not
change the DC conductance. This is a well-known result that has been
demonstrated from formally correct theories\cite{Maslov2} and which
is reproduced by our phenomenological approach.

We illustrate the physical consequences of the energy-induced
crossover from a transparent NS interface to the tunneling regime by
looking at the behavior of the differential resistance calculated
from Eq.~\ref{I-BTK}, that we integrate numerically after replacing
$z$ with $z_{e-e}$. In particular, we look at the dependence of the
differential resistance on temperature and voltage for different
values of $z$ and $\alpha$. The different $dV/dI$ curves are
normalized to $R_N = (1 + z^2)/G_0$, the normal state resistance in
the non-interacting case ($\alpha = 0$). Note that, in general and
contrary to the case of non-interacting electrons, the differential
resistance of the interacting system depends on the bias even when
$eV > \Delta$, because of the energy dependence of the transmission
probability.

Fig.~\ref{Temp-Dep} shows the temperature dependence of the linear
resistance of the NS interface for different values of the (bare)
$z$ parameter, enabling the direct comparison of the non-interacting
($\alpha = 0$) and the interacting case (with $\alpha = 0.15$ in
this example). As explained above, interactions do not have any
influence if $z=0$ and the temperature dependence of the linear
resistance is identical in the two cases. At finite values of $z$,
however, the effect of the interaction is visible. In particular, at
low $z$ ($z=0.22$) and in the absence of interaction the resistance
of the non-interacting system decreases with decreasing $T$ below
$T_c$, so that at low temperature the resistance is smaller than
that measured for $T>T_c$. However, in the presence of interactions
the behavior of the resistance is qualitatively different. After an
initial decrease (just below $T_c$), the resistance increases again
rapidly, so that at low temperature it is larger than in the normal
state.

This difference in behavior is a clear manifestation of the crossover
from the Andreev to the tunneling limit mentioned above. For the
interacting system, the thermal energy of the electrons at $T \simeq
T_c$ is sufficiently large to prevent a strong renormalization of
the transmission probability at the interface, and the observed
behavior (i.e., the decrease in resistance) is similar to the
non-interacting case in the Andreev limit. At lower temperature,
however, the renormalized transmission probability becomes small and
the interacting system goes into the tunneling limit. As a
consequence the resistance increases. The precise temperature at
which the upturn occurs depends on $z$ and on the interaction
strength $\alpha$, shifting to higher values with increasing these
parameters. As shown in the figure, for sufficiently large values of
$z$ the resistance increases with lowering $T$ across $T_c$ both in
the interacting and the non-interacting case, and the difference in
behavior becomes only of quantitative nature.

\begin{figure}[b]
  \centering
  \includegraphics[width=8.5cm]{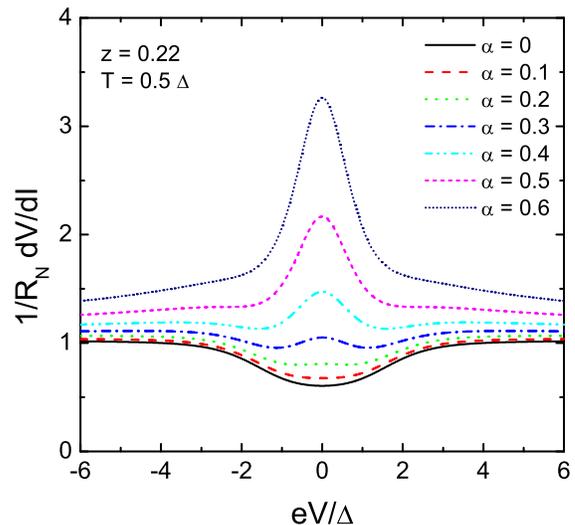}
  \caption{The $dV/dI-V$ calculated for a fixed interface transparency
   ($z=0.22$) and temperature $T=0.5\ \Delta$, as a function of interaction
    strength ($\alpha$ ranging from 0 to 0.6). Increasing the
    interaction results in the complete disappearance of the resistance
    suppression seen below the gap for the non-interacting system.}\label{Diff-as}
\end{figure}

A second manifestation of the crossover from Andreev to tunneling
behavior can be seen in the $dV/dI$-$V$ curves measured at different
temperatures. Also here, the case exhibiting qualitative differences
for the interacting and non-interacting cases is that of a small
(bare) $z$ parameter. The results of the calculations are shown in
Fig.~\ref{Diff-Ts}, where $z=0.03$ and $\alpha = 0.6$. At higher
temperature (approximately $0.3\ \Delta$ in the present case), a
pronounced broad minimum in $dV/dI$ is observed when the DC bias is
smaller than the superconducting gap, as it is characteristic for
Andreev reflection. As the temperature is lowered, however, a sharp
peak in the $dV/dI$ curves appears around zero bias, causing the
resistance to exceed the normal state value. This peak cannot be
accounted for by the BTK-theory for non-interacting systems. Again,
the $dV/dI$ peak is a consequence of the renormalization of the
transmission probability which brings the NS interface in the
tunneling regime at sufficiently low energy.

Finally, it is instructive to look at the dependence of the
$dV/dI$-$V$ curves as a function of the electron-electron
interaction strength $\alpha$. The results of the calculations are
shown in Fig.~\ref{Diff-as} for the case of $z=0.22$ and $T=0.5\
\Delta$, with $\alpha$ ranging from $0$ to $0.6$. In the
non-interacting case ($\alpha = 0$), transport occurs in the Andreev
regime for all values of the applied bias, resulting in a
suppression of the resistance at low energy. However, as the
interaction strength is enhanced, a peak in $dV/dI$ appears around
zero bias, which becomes broader for larger values of $\alpha$.
Eventually, this peak completely dominates the behavior of the
differential resistance for bias voltages below the superconducting
gap. In this case, the resistance suppression characteristic of
Andreev reflection is not visible any more. This is because, even
though the bare value of the transmission coefficient has remained
the same, the actual value of the transmission has changed due to
the energy dependent renormalization induced by electron-electron
interaction. In a non-interacting picture the observation of such a
peak in the $dV/dI$-$V$ curve would be tentatively interpreted as
due to a low transparency interface. One would however note that the
shape of the curve is very different from what is normally observed
in the tunneling regime, as indicated, for instance, by the absence
of any feature at a bias corresponding to the superconducting gap.

The temperature and bias voltage dependence of the differential
resistance that we find in the interacting case for high (bare)
values of the transmission through the NS interface reproduce
qualitatively what has been found previously from formally correct
calculations for a Superconductor-Luttinger Liquid
junction\cite{Lee}. This indicates that the phenomenological picture
presented here is physically sound. We want to explicitly make
clear, however, that a Luttinger liquid behavior of the normal metal
is not a necessary requirement for the observation of the behavior
that we have described. All is needed is a sufficiently rapid
suppression of the transmission probability with lowering energy on
the scale of the superconducting gap. As mentioned above, in
low-dimensionality samples such a suppression can be caused by
electron-electron interaction under fairly general circumstances.

Finally, our results exhibit qualitative agreement with the behavior
observed experimentally in carbon nanotubes contacted with
superconducting electrodes. The case of high transmission contacts
has been studied in Ref.~[12], where an anomalous (as compared to
the non interacting case) temperature and bias dependence of the
differential resistance has been reported. The experimental
observations are qualitatively very similar to the behavior shown in
Fig.~\ref{Temp-Dep} and \ref{Diff-Ts} for the low $z$ case. The case
of lower transmission contacts has been studied in Ref.~[16]. In
that experiment a tunneling-like BCS-density of state was expected
and observed without any clear "BCS shoulders". The absence is
attributed to the presence of interaction which decreases the
magnitude of this large feature. This behavior is similar to the one
predicted by our calculations, as illustrated in Fig.~\ref{Diff-as}
(curves with $\alpha =0.5$ and $0.6$).

In conclusion, we have proposed a phenomenological description of
the effect of electron-electron interaction on transport through a
normal conductor/superconductor interface of arbitrary transparency.
The results obtained from this description agree qualitatively with
what is predicted by formally correct theories and reproduce the
behavior observed experimentally in carbon nanotube/superconductor
junctions. We expect that the simplicity of the picture proposed
here will be useful in the interpretation of experimental data.


\begin{thebibliography}{99}
\bibitem{note}  In Eq.~\ref{T_of_E} $T(E)$ vanishes at $E=0$, whereas in different experimental systems
the interaction induced suppression of $T(E)$ can saturate at a finite value. Although it is important
to bear this point in mind, this does not change the qualitative conclusions of our work.
\end{thebibliography}
\end{document}